\documentclass[11pt]{article}
\usepackage{graphicx,amsmath,amssymb,epsf} \usepackage{latexsym,bm,
  slashed} 
  \usepackage{xcolor} 
  \definecolor{dark}{rgb}{0.10,0.2,0.3}
\definecolor{magenta}{rgb}{0.7,0.1,0.3}
\definecolor{purpure}{rgb}{0.5,0.15,0.3}
\usepackage[font=small,format=plain,labelfont=bf,up,textfont=it,up]{caption}
\usepackage{hyperref, cite} 
\hypersetup{colorlinks, citecolor=blue,
  filecolor=blue, linkcolor=magenta,
  urlcolor=purpure}

  \usepackage{xcolor}
\usepackage{ifthen}
\usepackage{amsmath}
\usepackage{amssymb}
\usepackage{mathtools, nccmath}
\usepackage{graphicx, wrapfig}
\usepackage{slashed, bm}
\usepackage[utf8]{inputenc}
\usepackage{authblk}
\usepackage[normalem]{ulem}

\newcommand{\kt}{{\bm k}}

\newcommand{\qt}{{\bm q}}
\newcommand{\Dt}{{\bm \Delta}}

\definecolor{kkcolor}{rgb}{0,0.6,0.4}

\definecolor{mhcolor}{rgb}{0.7,.1,0.7}

 \title{Evidence for the maximally entangled low $x$ proton  in Deep Inelastic Scattering from  H1 data
 }

\author[1]{Martin~Hentschinski}
\author[2]{Krzysztof~Kutak}

\affil[1]{\normalsize Departamento de Actuaria, F\'isica y Matem\'aticas, 
Universidad de las Americas Puebla, San Andr\'es Cholula, 72820 Puebla, Mexico }
\affil[2]{ \normalsize Institute of Nuclear Physics, Polish Academy of Sciences, 
 ul.~Radzikowskiego 152, 31-342, Krak\'ow, Poland}

\begin{document}

\maketitle
\begin{abstract}
We investigate the proposal by Kharzeev and Levin of a maximally entangled proton wave function in Deep Inelastic Scattering at low $x$ and the proposed relation between parton number and final state hadron multiplicity.   Contrary to the original formulation we determine partonic entropy from the sum of gluon and quark distribution functions at low $x$, which we obtain from an unintegrated gluon distribution subject to next-to-leading  order Balitsky-Fadin-Kuraev-Lipatov evolution. We find for this framework very good agreement with H1 data. We furthermore provide a comparison based on   NNPDF parton distribution functions at both next-to-next-to-leading order  and next-to-next-to-leading with small $x$ resummation,  where the latter provides an acceptable description of data.

\end{abstract}

\section{Entanglement entropy}
\label{sec:entanglement-entropy}
The proton is a coherent quantum state with zero von Neumann
entropy. However it has been argued in 
 \cite{Kharzeev:2017qzs,Tu:2019ouv}
that when the proton wave function is observed in Deep Inelastic Scattering (DIS) of
electrons and protons, this is no longer true. In DIS, the virtual photon, with momentum $q$ and $q^2 = - Q^2$ its
virtuality, probes only parts of the proton wave function, which
gives rise to entanglement entropy,  between observed and unobserved
parts of the proton wave function, through tracing out inaccessible
degrees of freedom of the density matrix. The resulting entanglement is then  a measure of the degree to which the probabilities in the two subsystems are correlated; for other approaches where thermodynamical and momentum space entanglement entropy have been studied see  \cite{Kutak:2011rb,Peschanski:2012cw,Kovner:2015hga,Kovner:2018rbf,Armesto:2019mna,Duan:2020jkz,Dvali:2021ooc,Ramos:2020kaj, Kharzeev:2021nzh}; for studies on Wehrl entropy \cite{Hagiwara:2017uaz} and jet entropy see \cite{Neill:2018uqw}).  Based on 
explicit studies of this entanglement entropy , both within a 1+1
dimensional toy model and leading order (LO) Balitsky-Kovchegov
evolution \cite{Kovchegov:1999yj,Kovchegov:1999ua,Balitsky:1995ub}, as well as entanglement entropy in conformal field theory, the authors of
\cite{Kharzeev:2017qzs} conclude that DIS probes in the perturbative low
$x$ limit a maximally entangled state.
 With $x =Q^2/2 p \cdot q$ and
$p$ the proton momentum, the low $x$ limit corresponds to the
perturbative high energy limit, where $Q^2$ defines the hard
scale of the reaction and  sets the  scale of the strong running coupling
constant $\alpha_s(Q^2) \ll 1$. The perturbative low $x$ limit of \cite{Kharzeev:2017qzs} corresponds then  to the 
 scenario where parton densities are
high, but not yet saturated and non-linear terms in the QCD evolution
equations are therefore sub-leading. This is precisely the kinematic
regime, where perturbative low $x$ evolution of the proton is
described through
Balitsky-Fadin-Kuraev-Lipatov (BFKL) evolution, which resums terms  $[\alpha_s \ln(1/x)]^n$ to all order in $\alpha_s$; it is this kinematic regime to which the results of \cite{Kharzeev:2017qzs} are supposed to apply at first. \\

The proposal that DIS probes in the low $x$ limit a maximally
entangled state is closely related to  the emergence of an exponentially large number of partonic micro-states which occur with equal probabilities $P_n(Y) = 1/\langle n\rangle$, with $\langle n (Y, Q)\rangle$  the average number of partons at  $Y = \ln 1/x$ and photon virtuality $Q$. Entropy  is then directly obtained as
\begin{align}
  \label{eq:vonNeuman}
  S(x, Q^2)& = \ln \left\langle n\left(\ln\frac{1}{x}, Q \right)\right\rangle.
\end{align}
Assuming that the second law of thermodynamics holds for this entanglement entropy, the above expressions yields a lower bound on the entropy of final states hadrons $S_h$ through $S_h \geq  S(x, Q^2)$ \cite{Kharzeev:2017qzs}.  ``Local parton-hadron duality'' \cite{Dokshitzer:1987nm} and the ``parton liberation'' picture  \cite{Mueller:1999fp} then suggest that  partonic entropy coincides with the entropy of final state hadrons in DIS, see also the discussion for hadron-hadron collisions in \cite{Kutak:2011rb}. The hadronic entropy can be further related to the multiplicity distribution of DIS final state hadrons. The latter has been obtained from HERA data in \cite{H1:2020zpd},  which allows for a direct comparison of Eq.~\eqref{eq:vonNeuman} to experimental data. \\

Confirmation of Eq.~\eqref{eq:vonNeuman} is of high interest, since it links hadron structure to final state multiplicities through entropy. If confirmed, it provides an additional constraint on parton distribution functions (PDFs). Moreover, entropy is defined non-perturbatively and the proposed relation is therefore not necessarily limited to perturbative events, unlike PDFs. Last but not least, entropy is subject to  quantum bounds \cite{Bekenstein:1980jp, Bousso:1999xy, Ryu:2006bv} and through Eq.~\eqref{eq:vonNeuman} such bounds translate directly on bounds on the  number of partons in the proton \cite{Kharzeev:2017qzs}. This is of particular interest for the search for a saturated gluon state commonly called the Color Glass Condensate at collider facilities such as the Large Hadron Collider and the   future Electron Ion Collider. 
\\

The explicit model calculations of \cite{Kharzeev:2017qzs} were based on  solutions of purely gluonic LO low $x$ evolution, where quarks appear only as a next-to-leading order (NLO) correction;   it is therefore natural to assume that at first  the total numbers of  partons agrees with the number of gluons. In the following we find that for the kinematic regime explored at HERA, quarks are indeed sub-leading, but nevertheless numerically relevant for a correct description of data. We therefore propose in this letter that the average number of partons in Eq.~\eqref{eq:vonNeuman} should be interpreted  as the sum of the number of all partonic degrees of freedom, {\it i.e.} of quarks and gluons.
\\

Our description is based on the  NLO BFKL fit \cite{Hentschinski:2012kr,Hentschinski:2013id} (HSS). Initial conditions of the HSS unintegrated gluon distribution have been fitted to HERA data on the proton structure function $F_2$ and  the HSS fit provides therefore a natural framework to verify the validity of Eq.~\eqref{eq:vonNeuman} and its conjectured relation to the final state hadron multiplicity.  Moreover, the HSS fit is directly subject to NLO BFKL evolution \cite{Fadin:1998py} and therefore provides a direct implementation of linear QCD low $x$ evolution.

\section{Results}
\label{sec:collinearSea}
To  compare the HSS unintegrated gluon distribution to data, we need to determine first PDFs,  
which  will yield the total number of partons through
\begin{align}
  \label{eq:gluon}
  \left\langle n\left(\ln\frac{1}{x}, Q \right)\right\rangle = xg(x, Q) + x\Sigma(x, Q),
\end{align}
where $g(x, \mu_F)$ ($\Sigma(x, Q)$) denotes the  gluon (seaquark) distribution function at the factorization scale $\mu_F$. To this end we use the Catani-Hautmann procedure  \cite{Catani:1994sq} for the determination of high energy resummed PDFs. At leading order, the prescription is straightforward for the gluon distribution function, which is obtained as
\begin{align}
  \label{eq:gluon_collinear}
  xg(x, \mu_F) & = \int_0^{\mu_F^2} d \kt^2 {\mathcal{F}}(x, \kt^2),
\end{align}
where  $\mu_F$ denotes the factorization scale which we identify for the current study with the photon virtuality $Q$, and ${\mathcal{F}}(x, \kt^2)$ the unintegrated gluon distribution, subject to BFKL evolution.  To obtain the seaquark distribution, we require a transverse momentum dependent splitting function \cite{Catani:1994sq} ,
\begin{align}
  \label{eq:splitting}
\tilde{P}_{qg}\left(z,\frac{\kt^2}{\Dt^2} \right) & = \frac{\alpha_s 2n_f}{2 \pi} T_F \frac{\Dt^2}{[\Dt^2 + z(1-z) \kt^2]^2} \left[z^2 + (1-z)^2 + 4 z^2 (1-z)^2 \frac{\kt^2}{\Dt^2} \right],
\end{align}
where $\kt$ denotes the gluon momentum and $\Dt = \qt -z\kt$ with $\qt$ the $t$-channel quark transverse momentum; $T_F=1/2$. Note that this splitting function reduces in the collinear limit $\kt \to 0$ to the conventional leading order DGLAP splitting function
$ {P}_{qg}(z)  =  \frac{\alpha_s 2 n_f}{2 \pi} T_F \left[z^2 + (1-z)^2 \right] $.
The integrated seaquark distribution is then obtained as \cite{Catani:1994sq}
\begin{align}
  \label{eq:XSea}
  x\Sigma(x, Q)& = \int_0^\infty \frac{d \Dt^2}{\Dt^2} \int_0^\infty d\kt^2 \int_0^1 dz \Theta\left(Q^2 - \frac{\Dt^2}{1-z} - z \kt^2  \right) \tilde{P}_{qg}\left(z,\frac{\kt^2}{\Dt^2} \right) \mathcal{F}(x, \kt^2).
\end{align}
Note that in \cite{Hentschinski:2017ayz, Hentschinski:2021lsh} a corresponding off-shell gluon-to-gluon splitting function has been determined. Within the current setup, this would allow in principle for the determination of the gluon distribution at next-to-leading order. The use of this splitting function for the determination of the  gluon distribution function at NLO has however not been worked out completely so far. Moreover, the HSS fit is  based on a leading order virtual photon impact factors, which suggests the use of the leading order prescription Eq.~\eqref{eq:gluon_collinear} also for this study. The HSS unintegrated gluon density  reads \cite{Chachamis:2015ona}
\begin{align}\label{eq:Gudg}
  \mathcal{F}\left(x, {\bm k}^2, Q\right) 
& = 
 \frac{1}{{\bm k}^2}\int \limits_{\frac{1}{2}-i\infty}^{\frac{1}{2} + i \infty}  \frac{d \gamma}{2 \pi i}   \; \; \hat{g}\left(x, \frac{Q^2}{Q_0^2}, \gamma \right) \, \, \left(\frac{{\bm k}^2}{Q_0^2} \right)^\gamma,
\end{align}
where  $\hat{g}$ is an
operator in $\gamma$ space,
\begin{align}
  \label{eq:23}
 \hat{g}\left(x, \frac{Q^2}{Q_0^2}  \gamma \right)
  & = 
 \frac{\mathcal{C}\cdot \Gamma(\delta - \gamma)} {\pi \Gamma(\delta)}  \; \cdot \; 
 \left(\frac{1}{x}\right)^{\chi\left(\gamma, Q, Q \right)} \,\cdot \notag
 \\ 
&  
  \Bigg\{1    + \frac{\bar{\alpha}_s^2\beta_0  \chi_0 \left(\gamma\right) }{8 N_c} \log{\left(\frac{1}{x}\right)} 
  \Bigg[- \psi \left(\delta-\gamma\right)
 +  \log \frac{{Q}^2}{Q_0^2} -  \partial_\gamma \Bigg]\Bigg\}\;, 
\end{align}
with $\bar{\alpha}_s = \alpha_s N_c/\pi$, $N_c$ the number of
colors and   $\chi(\gamma, Q, Q)$  the next-to-leading
logarithmic (NLL) BFKL kernel which includes a resummation of both collinear enhanced terms as well as  a 
resummation of  large terms proportional to the first coefficient of the QCD
beta function, see App.~\ref{sec:HSS} for  details. 
 Eq.~\eqref{eq:gluon_collinear} and Eq.~\eqref{eq:XSea} is now used to calculate through  Eq.~\eqref{eq:gluon} the partonic entropy Eq.~\eqref{eq:vonNeuman}; the result is then compared to  H1 data \cite{H1:2020zpd}. To calculate entropy for the H1 $Q^2$ bins, we employ the following averaging procedure,
\begin{equation}
    \bar S(x)_{Q_2^2,Q_1^2}=\ln\frac{1}{Q_2^2-Q_1^2} \int_{Q_1^2}^{Q_2^2}dQ^2\left[xg(x,Q^2)+x\Sigma(x,Q^2)\right].
\end{equation}
 The results of our study are shown in Fig.~\ref{fig:results}, where we evaluate all expressions for $n_f=4$ flavors. 
 \begin{figure}[t]
    \centering
    \parbox{0.49\textwidth}{\includegraphics[width=0.49\textwidth]{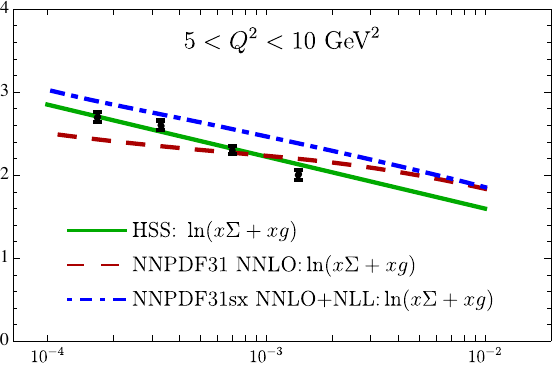}}$\,$ \parbox{0.49\textwidth}{\includegraphics[width=0.49\textwidth]{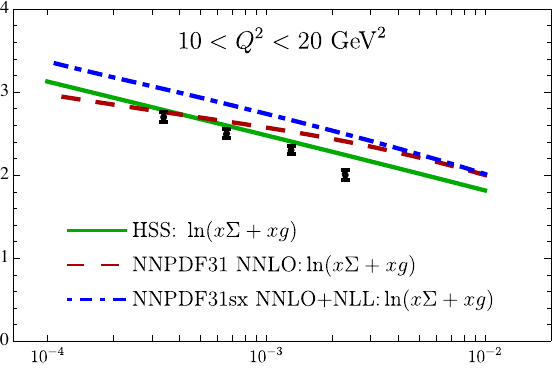}}
    \parbox{0.49\textwidth}{\includegraphics[width=0.49\textwidth]{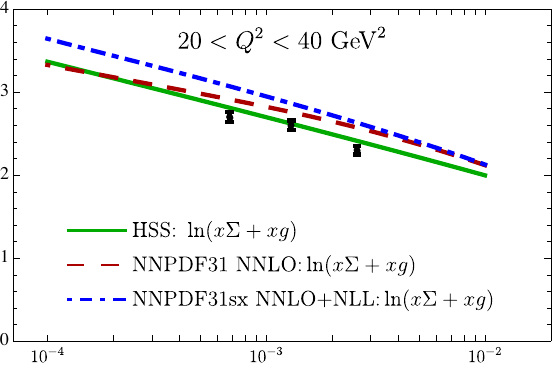}}
    \parbox{0.49\textwidth}{\includegraphics[width=0.49\textwidth]{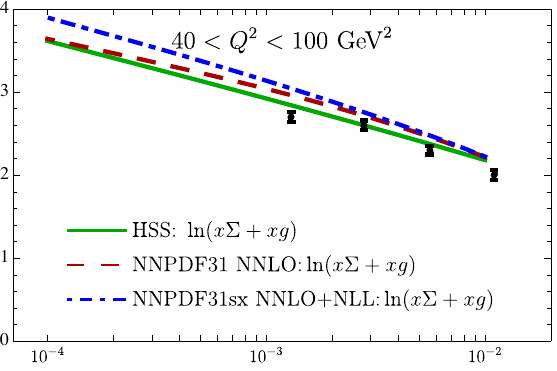}}
    \caption{Partonic entropy  versus Bjorken $x$, as given by Eq.~\eqref{eq:vonNeuman} and Eq.~\eqref{eq:gluon}. We furter show results based on the gluon distribution only  as well as a comparison to NNPDFs. Results are compared to the final state hadron entropy derived from the multiplicity distributions measured at H1 \cite{H1:2020zpd}  }
    \label{fig:results}
\end{figure}
We find that the partonic entropy obtained from the total number of partons gives a very good description of  H1 data  \cite{H1:2020zpd} in case of the HSS fit. As anticipated  in \cite{Kharzeev:2017qzs}, the purely gluonic contribution is clearly dominant and amounts to approximately 80\% of the total contribution; nevertheless the seaquark contribution is needed for an accurate description of H1 data. Given the approximations taken in the derivation of Eq.~\eqref{eq:gluon} as well as the possibility that sub-leading corrections are relevant for the determination of hadronic entropy form  the multiplicity distribution, we believe that the above result provides an impressive confirmation of Eq.~\eqref{eq:gluon} and the results of \cite{Kharzeev:2017qzs} in general. \\ 

In \cite{H1:2020zpd} the data shown in Fig.~\ref{fig:results} have been compared to Eqs.~\eqref{eq:vonNeuman} and \eqref{eq:gluon}. Based on the original proposal of \cite{Kharzeev:2017qzs}, only the gluon PDF has been used, for which the LO gluon distribution of the HERAPDF 2.0 set \cite{H1:2015ubc} has been chosen. While the use of a LO gluon PDF is somehow  natural, since Eq.~\eqref{eq:vonNeuman} does at the moment clearly not  address questions related to collinear factorization at NLO and beyond, it is  well known that the convergence of the gluon distribution is rather poor in the low $x$ region;  differences between the LO   and NLO gluon amount up to 100\% in the low $x$ region, see {\it e.g.} Fig.~26 of \cite{H1:2015ubc}. While there are still noticeable differences between NLO and NNLO gluon distribution (of the order of 30\% at $x=10^{-4}$), one can nevertheless argue that the gluon distribution starts to converge beyond leading order and the values provide by the NLO gluon might be taken as a more realistic reflection of the true gluon distribution. To substantiate this point, we show in Fig.~\ref{fig:results} also results based on an evaluation of Eqs.~\eqref{eq:vonNeuman} and \eqref{eq:gluon}   with NNPDF  collinear PDFs  at NNLO\cite{NNPDF:2014otw}. We further show results obtained using NNLO NNPDF with next-to-leading logarithmic (NLL) low $x$ resummation \cite{Ball:2017otu}. In both cases we assume   $\mu_F=Q$. While both PDF sets allow for an approximate description of data and may therefore serve as an additional confirmation of the correctness of Eq.~\eqref{eq:gluon},  a satisfactory description of the $x$-dependence is only possible using the low $x$ resummed NNPDF PDF set, which provides a very good description of the  shape, with a slight off-set in normalization.\\

A different description of these data has been provided in \cite{Kharzeev:2021yyf} which uses the sea quark distribution only. The authors use however for  their LO BFKL  description the   quark-to-gluon splitting function instead of the required gluon-to-quark splitting. The former is enhanced in the low $x$ limit and yields an incorrect sea quark distribution which is presumably of the order of the gluon distribution. We also could not reproduce  the description which is based on the  collinear NNLO sea quark distribution.  \\

\section{Conclusions}
\label{sec:concl}

In this letter we followed the proposal of \cite{Kharzeev:2017qzs} to treat the low $x$ proton in Deep Inelastic Scattering as a maximally entangled state with an entanglement entropy given as the logarithm of the average number of partons in the proton. Unlike \cite{Kharzeev:2017qzs, Kharzeev:2021yyf} we interpret the total number of partons as the sum of quarks and gluon numbers, determined through the regarding PDFs. While we agree with \cite{Kharzeev:2017qzs, Dvali:2021ooc, Kutak:2011rb}  that the quark distribution is sub-leading in the low $x$ limit, we find that the seaquark distribution provides a numerically relevant contribution of the order of 20\%. \\

Our description is  based  on the determination of  PDFs from an unintegrated low $x$ gluon distribution, subject to BFKL evolution. For the numerical study, the HSS unintegrated gluon, which follows NLO BFKL evolution, has been used.  Comparing our result with the final state hadron entropy extracted by the H1 collaboration \cite{H1:2020zpd}, we find a very good agreement with data,  if the total number of partons is taken as the sum of gluons and sea-quarks. We also provided a comparison based on NNLO PDF sets by the  NNPDF collaboration. While purely NNLO DGLAP PDFs provide only an approximate description of data, we find that  NNLO DGLAP PDFs with  NLL low $x$ resummation provide a  reasonable description of the slope of H1 data, which emphasis again the  role of low $x$ dynamics for the determination of the proton as a maximally entangled state of partons.  \\

Note that such an agreement is not obtained if the comparison is based on  leading order collinear PDFs, as  used by the H1 collaboration. This clearly hints at the need to further refine the underlying theoretical framework, in particular to clarify in a systematic way the relation between entropy and PDFs within the framework of collinear and/or high energy factorization.
This need is immediately apparent if Eq.~\eqref{eq:gluon} is evaluated using PDFs beyond leading order, which immediately implies a scheme dependence of the extracted parton number; strictly speaking the latter can be therefore no longer related to the hadron multiplicity which is a physical observable and therefore scheme independent. The description based on NNLO PDFs and NNLO low x resummed PDFs is therefore an approximation at best. Note that a similar limitation does not apply to the description based on the HSS fit, since the resulting PDFs are leading order, from the point of view of collinear factorization and therefore scheme dependent, while similar issues arise due to the use of high energy factorization beyond leading order in that case. Moreover
 the relation  to other frameworks as studied in \cite{Kutak:2011rb,Peschanski:2012cw,Kovner:2015hga,Kovner:2018rbf,Armesto:2019mna,Duan:2020jkz,Dvali:2021ooc} needs to be clarified.  Furthermore it will be interesting to  explore possible deviations from this framework at lower values of $Q$ and $x$ due to the onset of nonlinear low $x$ evolution,  in particular effects due to saturated  parton densities  \cite{Gribov:1983ivg,McLerran:1993ni}.

\section*{Acknowledgments}
\label{sec:Ack}
We would like to thank Stefan Schmitt for useful correspondence. 
KK acknowledges partial support by the Polish National Science Center with grant no. DEC-2017/27/B/ST2/01985.
MH  is grateful for support by
Consejo Nacional de Ciencia y Tecnolog\'ia grant number A1 S-43940
(CONACYT-SEP Ciencias B\'asicas).

\appendix

\section{Some details on the HSS NLO BFKL fit}
\label{sec:HSS}

The  NLL kernel with collinear
improvements which underlies the the NLO BFKL fit \cite{Hentschinski:2012kr,Hentschinski:2013id, Bautista:2016xnp} reads
 \begin{align}\label{eq:gluongf}
\chi\left(\gamma, M, \overline{M} \right) &=
{\bar\alpha}_s\chi_0\left(\gamma\right) +
{\bar\alpha}_s^2\tilde{\chi}_1\left(\gamma\right)-\frac{1}{2}{\bar\alpha}_s^2
\chi_0^{\prime}\left(\gamma\right)\chi_0\left(\gamma\right) +
\notag \\
&\hspace{2cm} 
+ \chi_{\text{RG}}({\bar\alpha}_s,\gamma,\tilde{a},\tilde{b}) - \frac{\bar{\alpha}_s^2}{8 N_c} \chi_0(\gamma) \log \frac{\overline{M}^2}{M^2}  .
\end{align}
where $\chi_i$, $i=0,1$ denotes the LO and NLO BFKL eigenvalue and
$\chi_{\text{RG}}$ resums (anti-)collinear poles to all orders;  see
\cite{Hentschinski:2012kr} for details.  The scale $M$ is a
characteristic hard scale of the process, while $\overline{M}$ sets
the scale of the running coupling constant. For the current study we set $M = \overline{M} = Q$ and $n_f=4$ with $\Lambda_{\text{QCD}} = 0.21$~GeV.
$Q_0 = 0.28$~GeV, and $\delta = 6.5$. have been determined from a fit to  the $F_2$ structure function in \cite{Hentschinski:2012kr}. In this fit the  overall running coupling constant  has been evaluated at the renormalization scale $\mu^2=QQ_0$, with $Q$ the photon virtuality. For the construction of parton distribution function $\mu^2=Q^2$  is however a more natural choice. We therefore reevaluated the underlying fit and found that data on the proton structure $F_2$ \cite{H1:2009pze}  are equally well described, if we use  $\mu^2=Q^2$ for the photon impact factor with a  normalization  $\mathcal{C}=4.31$. It is then this convention which we  use in this study.

\newpage
\begin{center}
  {\bf Erratum }
\end{center}

\begin{figure}[h!]
    \centering
    \parbox{0.49\textwidth}{\includegraphics[width=0.49\textwidth]{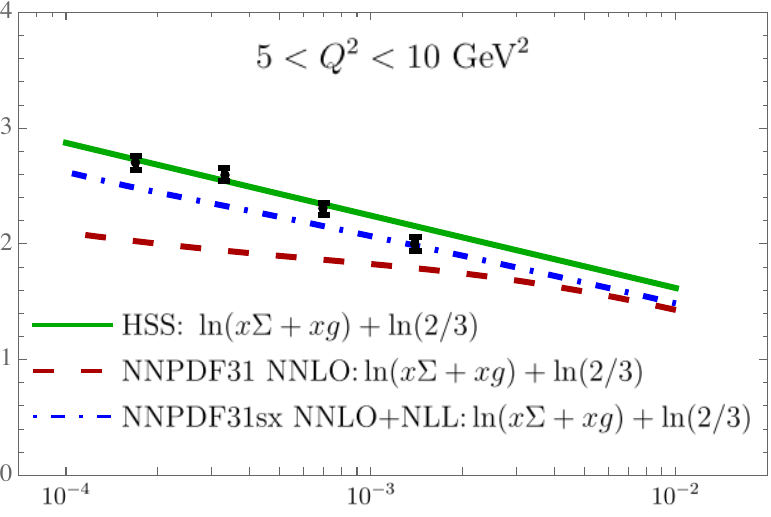}}$\,$ \parbox{0.49\textwidth}{\includegraphics[width=0.49\textwidth]{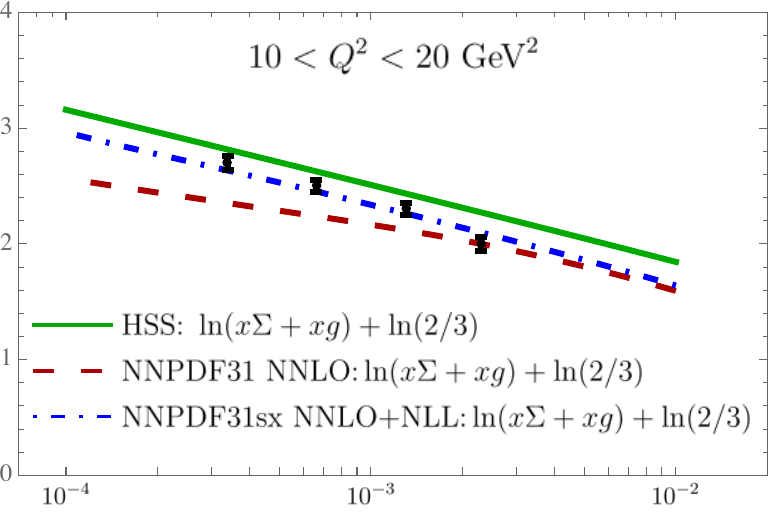}}
    \parbox{0.49\textwidth}{\includegraphics[width=0.49\textwidth]{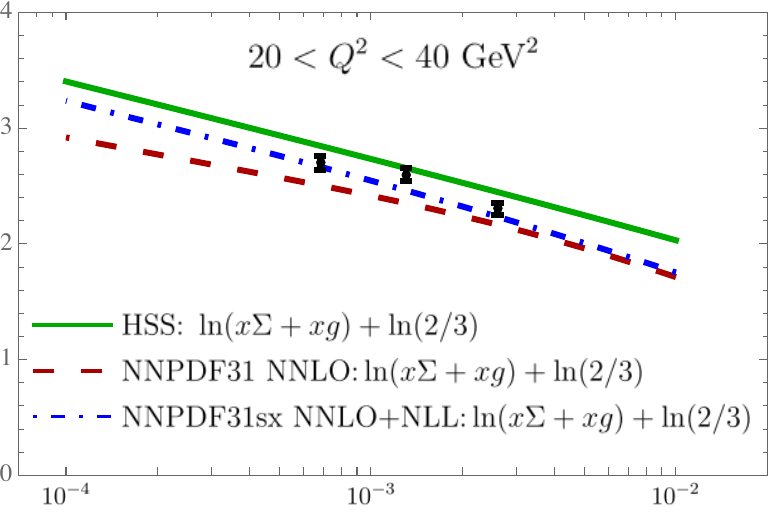}}
    \parbox{0.49\textwidth}{\includegraphics[width=0.49\textwidth]{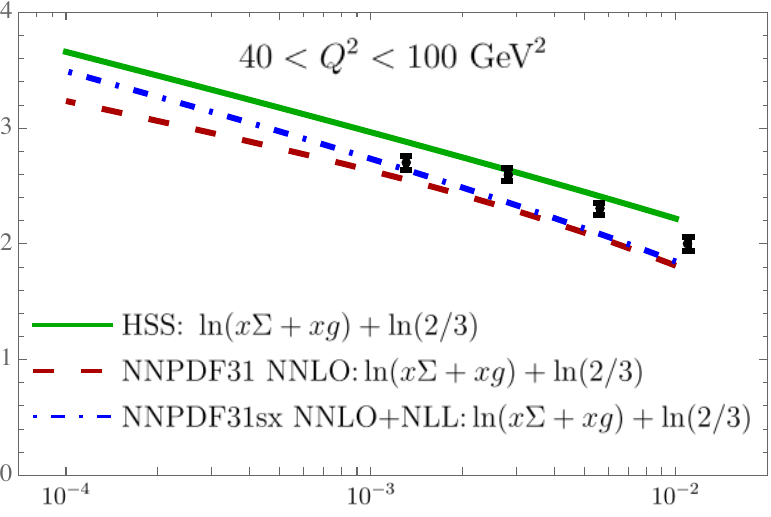}}
    \caption{Partonic entropy  versus Bjorken $x$, as given by Eq.~\eqref{eq:entropy} and Eq.~\eqref{eq:gluon}. We furter show results based on the gluon distribution only  as well as on quarks and gluons together. Results are compared to the final state hadron entropy derived from the multiplicity distributions measured at H1 \cite{H1:2020zpd}  }
    \label{fig:results}
\end{figure}

There was a  mistake in the scale choice of the running coupling in the gluon density that was was used in the paper \cite{Hentschinski:2021aux}; the mistake has been already  corrected in \cite{Hentschinski:2022rsa}.  The mistake was  difficult to spot  since the formulas that we used did not account for the fact that only the charged hadrons were measured. The numerical factors approximately  canceled  and the net result is only slightly changed. 
The number of partons in the  corrected formulas is \cite{Hentschinski:2022rsa}:
\label{sec:collinearSea}
\begin{align}
  \label{eq:gluon}
  \left\langle n\left(\ln\frac{1}{x}, Q \right)\right\rangle = \frac{2}{3}\left[xg(x, Q) + x\Sigma(x, Q)\right],
\end{align}
where $g(x, \mu_F)$ ($\Sigma(x, Q)$) denotes the  gluon (seaquark) distribution function at the factorization scale $\mu_F$ and as described above the factor $2/3$ takes into account the fact that only charged partons were measured, see also the more detailed discussion in \cite{Hentschinski:2022rsa}.  To calculate entropy for the H1 $Q^2$ bins, we employ the following averaging procedure,
\begin{equation}
 \label{eq:entropy}
    \bar S(x)_{Q_2^2,Q_1^2}=\ln\frac{1}{Q_2^2-Q_1^2} \int_{Q_1^2}^{Q_2^2}dQ^2 \left\langle n\left(\ln\frac{1}{x}, Q \right)\right\rangle.
\end{equation}
 The corrected results are shown in Fig.~\ref{fig:results}.\\
There was  also a typo in  Eq.~(5) of  \cite{Hentschinski:2021aux} which yields the formula for our determination of the 
sea quark distribution. The argument of the unintegrated gluon distriubtion has been given as $x$ instead of $x/z$  on the LHS of this equation.
The corrected formula, which was actually used in the calculation reads
\begin{align}
  \label{eq:XSea}
  x\Sigma(x, Q)& = \int_0^\infty \frac{d \Dt^2}{\Dt^2} \int_0^\infty d\kt^2 \int_0^1 dz \Theta\left(Q^2 - \frac{\Dt^2}{1-z} - z \kt^2  \right) \tilde{P}_{qg}\left(z,\frac{\kt^2}{\Dt^2} \right) \mathcal{F}\left(\frac{x}{z}, \kt^2\right).
\end{align}

\section*{Acknowledgments}
\label{sec:Ack}

MH  is grateful for support by
Consejo Nacional de Ciencia y Tecnolog\'ia grant number A1 S-43940
(CONACYT-SEP Ciencias B\'asicas).
KK acknowledges
the European Union’s Horizon 2020 research and innovation programme under grant agreement No. 824093.

\appendix

\end{document}